\documentclass{PoS}

\usepackage{graphics}
\usepackage{fancyhdr}
\usepackage{floatrow}
\usepackage{footmisc}

\usepackage[english]{babel}

\title{Neutron Position Sensitive Detectors for the ESS}

\ShortTitle{Neutron Detectors for ESS}

\author{\speaker{Oliver Kirstein}~$^{a,b}$,
Richard Hall-Wilton$^{a,c}$, Irina Stefanescu$^a$, Maddi Etxegarai$^a$,
Michail Anastasopoulos$^a$, Kevin Fissum$^{a,d}$ Anna Gulyachkina$^a$, Carina H\"oglund$^{a,e}$, Mewlude Imam$^{a,e}$, Kalliopi Kanaki$^a$, Anton Khaplanov$^a$, Thomas Kittelmann$^a$, Scott Kolya$^a$, Bj\"orn Nilsson$^{a,f}$, Luis Ortega$^a$, Dorothea Pfeiffer$^{a,g}$ Francesco Piscitelli$^{a,h}$, Judith Freita Ramos$^a$, Linda Robinson$^{a,e}$, Julius Scherzinger$^{a,d}$\\
\llap{$^a$}European Spallation Source, Lund, Sweden, \ 
\ \llap{$^b$}University of Newcastle, Callaghan, Australia\\
\llap{$^c$}Mid-Sweden University, Sundsvall, Sweden, \ 
\ \llap{$^d$}Lund University, Lund, Sweden\\
\llap{$^e$}Link\"oping University, Link\"oping, Sweden, \ 
\ \llap{$^f$}MAX-IV Laboratory, Lund, Sweden\\
\llap{$^g$}CERN, Geneva, Switzerland, \  
\ \llap{$^h$}ILL, Grenoble, France\\
E-mail: \email{oliver.kirstein@esss.se}\\
}

\abstract{
The European Spallation Source (ESS) in Lund, Sweden will become the world's leading neutron source for the study of materials. 
It will be a long pulse source, with an average beam power of 5 MW delivered to the target station. The ESS is in the construction phase, which started in 2013 with the completion of the Technical Design Report (TDR). 
The instruments are being selected from conceptual proposals submitted by groups from around Europe. These instruments present numerous challenges for detector technology in the absence of the availability of Helium-3, which is the default choice for detectors for instruments built until today and due to the extreme rates expected across the ESS instrument suite. Additionally a new generation of source requires a new generation of detector technologies to fully exploit the opportunities that this source provides. To meet this challenge at a green-field site, the detectors will be sourced from partners across Europe through numerous in-kind arrangements; a process that is somewhat novel for the neutron scattering community.
This contribution presents briefly the current status of detectors for the ESS, and outlines the timeline to completion.
For a conjectured instrument suite based upon instruments recommended for construction, a recently updated snapshot of the current expected detector requirements is presented. 
A strategy outline as to how these requirements might be tackled by novel detector developments is shown. In terms of future developments for the neutron community, synergies should be sought with other disciples, as recognized by various recent initiatives in Europe, in the context of the fundamentally multi-disciplinary nature of detectors. This strategy has at its basis the in-kind and collaborative partnerships necessary to be able to produce optimally performant detectors that allow the ESS instruments to be world-leading. This foresees and encourages a high level of collaboration and interdependence at its core, and rather than each group being all-rounders in every technology, the further development of centres of excellence across Europe for particular technologies and niches.
}

\FullConference{The 23rd International Workshop on Vertex Detectors,\\
		15-19 September 2014\\
		Macha Lake, The Czech Republic}

\begin{document}

\section{Preamble}

The European Spallation Source (ESS)~\cite{ESS} is currently under construction in Lund, Sweden, and foreseen to start a user programme early in the next decade.
The ESS  aspires to become the world's leading neutron source for the studies of materials during the next decade. 
Early scientific success during initial operations is an integral part of this vision. 
The ESS is jointly hosted by Sweden and Denmark, and is a Europe-wide project, presently comprising 17 partner countries, with discussions presently ongoing with several additional potential future partners. 
A large part of the funding will be available through in-kind contributions, meaning that there is significant of partners across Europe in the ESS construction. 
Given the `green-field' nature of the project with previously no extant host institute, the close involvement of partners and expertise from across Europe through in-kind and collaborative relationships is vital for the projects success.  

Given the unavailability and high price of $^3$He~\cite{cho,zeitelhack2012,PERSONS2011,KRAMER2011,kouzes3,SHEA2010} since 2009, detectors are perhaps the most critical technological challenge for the facility to address. 
This is often termed the `Helium-3 Crisis', however, as this has now been ongoing since 2009, it would be better named the `Helium-3 Reality'.
This message is further reinforced by the fact that latest predictions indicate that there will be no US supply after 2023~\cite{kouzes2013}, and the fact that stockpiles of Helium-3 are being rapidly diminished~\cite{russina2014}. 
For ESS, this means that replacing as much of the Helium-3 demand for the baseline initial instruments by alternate technologies is a necessity, as well as ensuring that alternatives exist to further reduce Helium-3 needs in the future in what will be an even sparser supply landscape for this rare gas~\cite{CDR,TDR,COLLABORATION-ICND}. 

This document presents a vision for the detectors and ongoing developments to ensure that ESS instruments have the detectors available that they require to be world-leading, and to ensure that they are available and highly-performant during early operations

\section{ESS instruments}

The ESS instruments are selected according to an open reviewed process of designs submitted by consortia from across Europe. 
Both the European scientific community and existing European neutron sources have input into this process as well as the ESS governance bodies. 
In order to determine the optimal choices for instruments, the following factors are considered, amongst others: 
\begin{itemize}
\item Strength of the future science case fulfilled by the instrument
\item Size of the user base for the instrument to ensure competitive utilisation
\item Quality of the conceptual design for the instrument and demonstration that it will be world leading
\item Engineering and technical reality of the achievability of the design. 
\item {\it ``Early Success Strategy'':} How quickly the instrument will start to produce scientific output.
\end{itemize}

Initially foreseen as 22 `public' instruments as baseline for the ESS construction project.
`Public'' implies that the instruments are available for scientific users to bid for research time, through a reviewed proposal system.  
In the TDR~\cite{TDR}, a reference suite is outlined, which is one set of choices which could match the future scientific needs of European users. 

Presently 3 instruments are chosen and under construction since 1 January 2014. 
Another 9 instruments have been recommended for construction by the ESS scientific advisory committee, and subject to final approval, will start construction during 2015-16. 
This consists half of the baseline instruments for the ESS construction project. 

The instruments chosen fit into the following instrument classes depending upon their intended application area: 
\begin{itemize}
\item Neutron Imaging: {\it ODIN}
\item Small Angle Neutron Scattering (SANS): {\it LOKI, SKADI}
\item Neutron Macromolecular Crystallography: {\it NMX}
\item Diffraction: {\it DREAMS, HEIMDAL}
\item Engineering Diffraction:  {\it BEER}
\item Reflectometry:  {\it ESTIA, FREIA}
\item Spectroscopy:  {\it CSPEC, VOR, CAMEA}
\end{itemize}

\begin{figure}[htbp]
\begin{center}
\includegraphics[width=1.0\textwidth]{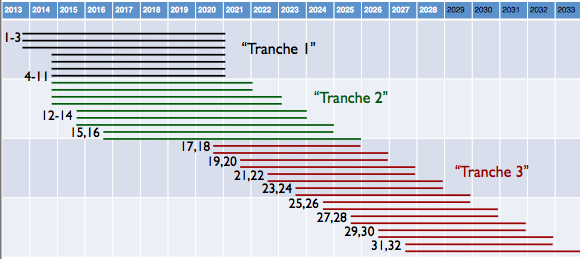}
\caption{\textbf Indicative diagram of the construction schedule for the ESS public instruments.}
\label{figure:instrument_schedule}
\end{center}
\end{figure}

An example timeline is shown in figure~\ref{figure:instrument_schedule} for the construction schedule of instruments. 
The details of this schedule will vary depending up the details of each instrument. 

The rest of this contribution will focus on how the detectors for these instruments might look. 

\section{Neutron Detector Technologies for ESS}

Neutrons are ``detected'' by a nuclear interaction in a ``converter material'' which leads to the destruction of the neutron in the process. The daughter products of this nuclear interaction are then characterised electronically by the detector. 
Most flagship instruments at current neutron sources use $^3$He gas as this converter material~\cite{ILL-blue-book}.  
New kinds of neutron detectors, not based on $^3$He, are urgently needed, due to the very limited availability of $^3$He. For this reason the detector group was the first neutron technology group established in July 2010, as well as extensive collaboration with European collaborative and in-kind partners on novel detector technologies during the ESS design update phase between 2010 and 2013. 

The result of this detector development programme is that a range of advanced prototypes and technology demonstrators exist for different application niches for particular classes of instruments. 
However, it is also clear that for ESS instruments, that the detector technologies will less monochrome than for existing sources~\cite{TDR,hall-wilton2012,MPGD2013}, where to a great extent only Helium-3 and a smaller fraction of $^6$Li-based scintillators are utilised. 
For ESS, these will be detectors based upon:

a) $^{10}$Boron Carbide thin film based gaseous detectors~\cite{BF3-B10-Workshop,hoglund2012}. These can be in a perpendicular-\cite{KLEIN,CASCADE,CORREA,KHAPLANOV2012,BIGAULT2012,ANDERSEN2012,CORREA2013,STEFANESCU2013a,STEFANESCU2013b} or parallel-~\cite{HENSKE2012,PISCITELLI,PISCITELLI2014,KANAKI2013,NOWAK2014} neutron incidence geometry. 

b)  $^6$Li-based scintillator detectors - both those utilising $^6$Li-glass with a coupled- or anger- style readout~\cite{HEIDERICH1991,KEMMERLING2001} and those incorporating $^6$Li embedded in a ZnS scintillating matrix, coupled to wavelength-shifting fibres~\cite{NAKAMURA2009,NAKAMURA2012,POWGEN,NIGEM}. 

c) For smaller, and lower rate applications, $^3$Helium detectors may be used.

d) Gd-based detectors, where resolution is needed, and gamma rejection can be relaxed~\cite{Schulz1999}.

To be consistent with the ``early success strategy'', it is important that ESS instruments are performant immediately upon the the provision of first neutrons. 
As such, it is vital that all installed detectors can be verified to be working and the developments are at such a stage of maturity to ensure this. 
This can be very simply summarized as follows: 

\begin{center}
{\it No prototypes installed in ESS}
\end{center}

In terms of reliability, detector systems should be designed with a 10 year minimum instrument lifetime in mind, with a view to being usable for an extended timeframe. 
The detector systems should be easily maintainable throughout that period, at a reasonable and low level of expected maintenance and expert intervention. 
The detector systems should be easily serviceable, so that repair and maintenance can be carried out as needed during regular non-beam periods.
The complexity of any service or regular maintenance required must be minimal to allow any trained detector technician to perform it; it should not require detailed knowledge of that particular system. 

The detector systems should be transparent to users and require no user-interventions. 
As such they must appear as the proverbial `black box' system. 
In contrast detailed information must be available to experts to allow them to verify the correct functioning of all components. 
All calibration must be transparent to users. 
Such calibration as needed must either be done by either by experts on a very irregular basis (yearly), or fully automated. 
Such requirements are analogous to those for photon detectors in the light-source community. 

\section{Detectors for ESS Instruments}

This section presents one set of choices for detectors which satisfy the requirements of the instruments recommended for construction at ESS so far. 
Note that the choices outlined here are not unique, but rather one possible set of possible choices to match the requirements of these instruments. 
As the maturity of these instrument designs develops over the coming years, during the conceptual and detailed design phases, the choices in detector technology will become clearer, and in some cases may even change. 

The detector requirements for instruments as presented here in table~\ref{DETREQ} are an update upon those presented last year ago in the TDR, and 2 years ago in the CDR, and now related to specific instrument designs recommended for construction rather than a generic reference suite. 

\begin{table}[thb] \footnotesize
\centering
 \begin{tabular}{lcccc} 
\hline
    {\bf Instrument}&{\bf Detector}& {\bf Wavelength} & {\bf Time}  &  {\bf Spatial} \\
  & {\bf area} & {\bf range} & {\bf resolution} & {\bf resolution} \\
   &[m$^2$]& [\AA]  & [$\mu$s] &   [mm] \\
\hline
Multi-Purpose Imaging\footnote{Tranche 1\label{n1}}  (ODIN) & 1 & 1 - 10      & 1        & 0.001 - 1  \\
&&&&\\
Broad-Band Small Sample SANS\footref{n1} (LOKI)                  & [8 - 16] & 3 - 20      & 100 & 2 - 8\\ %
General Purpose Polarised SANS\footnote{Tranche 2\label{n2}}  (SKADI)                & 2.0674  &  2 - 18 & 100 & 5 - 10\\ %
&&&&\\
Horizontal Reflectometer\footref{n2}   (FREIA)                                      & 0.25& 2 - 23      & 100 & 1$\times$8\\ %
Vertical Reflectometer\footref{n1}   (ESTIA)                        & 0.16& 5 - 9.4  & 100 & 0.5$\times$2\\ %
&&&&\\
Bi-Spectral Powder Diffractometer\footref{n1}   (POWTEX)                     &11.69 & 0.5 - 20  & $<$~10 & 2~$\times$~2 \\ %
Thermal Powder Diffractometer\footref{n2}  (HEIMDAL)        & 15.002 & 1 - 13  & 100 & 3~$\times$~3 \\%
&&&&\\
Material Science \& Engineering Diffractom.\footref{n1}   (BEER)  & 6.4925 & 0.1 - 7  & 10 & 2~$\times$~5  \\ %
&&&&\\
Macromolecular Diffractometer\footref{n1}  (NMX)                         & 1.08   & 1.8 - 3.5  & 1000 & 0.2 \\ %
&&&&\\
Cold Chopper Spectrometer\footref{n1}  (C-SPEC)                & 47.47    & 1.5 - 20  & 10 & 25~$\times$~25\\    %
Bi-Spectral Chopper Spectrometer\footref{n1}   (VOR)                   & 25.65  & 0.8 - 20  & 10      & 20~$\times$~20 \\ %
&&&&\\
Inverse TOF Spectrometer\footref{n2}   (CAMEA)            & 2.4   & 1 - 8         & $<$~10 & 5~$\times$~5\\%
&&&&\\

{\bf Total} &{\bf  [130 - 138]} &&& \\

\hline
\end{tabular}
\caption[Estimated detector requirements for the reference suite.]{Estimated detector requirements for the 22 reference instruments in terms of detector area, typical wavelength range of measurements and desired spatial and time resolution. The foot notes indicates the tranche in which the instruments is presently intended to be delivered. }
\label{DETREQ}
\end{table}


Using the information from the recommened instruments in table~\ref{DETREQ}, and combining this with those from other proposals which were not recommended, as well as those which are presently being designed, it is possible to generalise these detector requirements according to instrument class. 
These consolidated detector requirements grouped against instrument class is shown in table~\ref{CLASSREQ}.

\begin{table}[thb] \footnotesize
\centering
 \begin{tabular}{lccccc} 
\hline
    {\bf Instrument}& {\bf Wavelength} & {\bf Time}  &  {\bf Detector}&{\bf Spatial} &{\bf Rate}\\
  &  {\bf range} & {\bf resolution} &{\bf area} & {\bf resolution} &{\bf sample}\\
   &[\AA]  & [$\mu$s] &   [m$^2$]& [mm] &[n/s/cm$^2$]\\
\hline
SANS   & 3 - 20 & 100 [$\mu$s] & [10 - 18]&5&10$^{9}$\\
REFL   & 2 - 23 & 100 [$\mu$s] & 0.41&0.5&10$^{9}$\\
DIRECT   & 0.8 - 20 & 10 [$\mu$s] &73.12&  10 - 20 &10$^{7}$\\
INDIRECT   & 1 - 8 & [$\mu$s] & 2.4& 5& 10$^{10}$\\
DIFF   &0.5 - 20 & 10-100 [$\mu$s] & 26.692&2 - 10&10$^{9}$\\
NMX   & 1.8 - 3.5& [ms] & 1.08&0.2&10$^{8}$\\
IMAGING   &1 - 10 & [$\mu$s] &   1& 0.014-1&10$^{8}$\\
ENG   &0.1 - 7 & 10 [$\mu$s] & 6.4925& 5 &10$^{7}$\\
\hline

\hline
\end{tabular}
\caption{Estimated detector requirements for each instrument class in terms of typical wavelength range of measurements, detector area, desired spatial and time resolution and neutron rate on the per cm$^2$ on the sample.}
\label{CLASSREQ}
\end{table}

It is of course apt to take the detector requirements as outlined in tables~\ref{DETREQ} and~\ref{CLASSREQ} and try and prescribe appropriate technology choices for the detectors for each instrument, where these requirements might be met by those particular detector technologies.
It is important to consider the level of development of those technologies at the moment, and the time at which the instrument needs to be operational when considering this. 
Such an evaluation is shown in table~\ref{DETTECH}.

\begin{table}[!h] \footnotesize
\centering
\begin{tabular}{l|cc|cc|c|cc}  
\hline
 {\bf Instrument}&\multicolumn{7}{c}{\bf Detector Technology}\\
 & \multicolumn{2}{|c|}{\bf $^{10}$B Thin Films} &   \multicolumn{2}{|c|}{\bf Scintillators} & {\bf $^3$ He}  & \multicolumn{2}{|c}{\bf Exotica} \\ 
 & $\perp$ & $\parallel$ & WLS & Anger & & Gd & Other\\
\hline
ODIN& - & -  & - & o & - & o & +   \\    %
&&&&&&&\\
LOKI& o & +  & - & o & - & - & -   \\    %
SKADI& o & o  & - & + & - & - & -   \\    %
&&&&&&&\\
FREIA& - & +  & o & o & + & - & -   \\    %
ESTIA& - & +  & o & o & + & - & -   \\    %
&&&&&&&\\
HEIMDAL& o & o  & + & - & - & - & -   \\    %
DREAMS& o & +  & o & - & - & - & -   \\    %
BEER& + & +  & o & - & - & - & -   \\    %
&&&&&&&\\
NMX & o & o  & o & o & - & + & o   \\    %
&&&&&&&\\
C-SPEC& + & -  & - & - & - & - & -   \\    %
VOR& + & -  & - & - & - & - & -   \\    %
&&&&&&&\\
CAMEA & + & -  & - & - & + & - & o   \\    %
\hline
\end{tabular}
\caption{Appropriate detector technology options for the recommended instruments.  The detector technologies are grouped into perpendicular ($\perp$)- and inclined ($\parallel$)- neutron incidence geometries for $^{10}$B thin film detectors, wavelength shifting fibers (WLS) and Anger/direct-coupled cameras for scintillator detectors, $^3$He detectors, Gd-based detectors and Other. 
In the matrix of options, `+' indicates that this technology is presently seen as a high possibility, `-' indicates that it is a disadvantageous technology for this instrument, and `o' means that it is considered an option, though not the primary one.
}
\label{DETTECH}
\end{table}

\section{Conclusion}

With the ESS construction underway, the requirements for detectors for ESS instruments is becoming clearer. 
With significant developments for Helium-3 replacement technologies for detectors there will be a rich and complete suite of detectors to fulfill the needs of ESS instruments; detectors for ESS instruments are possible, despite the Helium-3 crisis. 
Note that as the requirements will still significantly evolve with the instrument design over the coming 1-2 years, these choices need to be continually reviewed. 

As with any crisis, it is not only about challenges, but also opportunities are opened up; with the new Helium-3 reality, the neutron scattering community is presented with an opportunity to enhance the possibilities for neutron instrumentation, in particular detectors, beyond that previously possible.


\end{document}